\documentclass{article}
\usepackage{spconf,amsmath,graphicx,hyperref}
\usepackage{booktabs}
\usepackage{multirow}
\usepackage{makecell}

\usepackage{cuted} 
\usepackage{subcaption}

\usepackage{caption}
\usepackage[most]{tcolorbox}
\usepackage{lipsum} 

\newtcolorbox[auto counter, list inside=cases]{casebox}[2]{
    colback=black!5,          
    colframe=black!75,        
    fonttitle=\bfseries,      
    title={Real case study \thetcbcounter: #1}, 
    label={#2},               
    arc=0mm,
    boxrule=1pt,
    left=4mm, right=4mm, top=2mm, bottom=2mm,
    attach boxed title to top left={yshift=-0.25mm-\tcboxedtitleheight/2, xshift=10mm},
    boxed title style={
        arc=3mm,
        colback=white,
        colframe=black!75,
        boxrule=0.5pt,
        left=2mm, right=2mm,
    }
}

\title{Revisiting Backdoor Threat in Federated Instruction Tuning from a Signal Aggregation Perspective}
\name{Haodong Zhao$^1$\qquad Jinming Hu$^1$\qquad Gongshen Liu$^{1,2}$\thanks{This work was sponsored by the Joint Funds of the National Natural Science Foundation of China (Grant No.U21B2020) and Science and Technology Cooperation Program of Shanghai Jiao Tong University in Inner Mongolia Autonomous Region—Action Plan of Shanghai Jiao Tong University for ``Revitalizing Inner Mongolia through Science and Technology''.}}
 
\address{$^1$ School of Computer Science, Shanghai Jiao Tong University\\
$^2$ Inner Mongolia Research Institute, Shanghai Jiao Tong University}
\begin{document}
\ninept
\maketitle
\begin{abstract}
Federated learning security research has predominantly focused on backdoor threats from a minority of malicious clients that intentionally corrupt model updates. This paper challenges this paradigm by investigating a more pervasive and insidious threat: \textit{backdoor vulnerabilities from low-concentration poisoned data distributed across the datasets of benign clients.} This scenario is increasingly common in federated instruction tuning for language models, which often rely on unverified third-party and crowd-sourced data. We analyze two forms of backdoor data through real cases: 1) \textit{natural trigger (inherent features as implicit triggers)}; 2) \textit{adversary-injected trigger}.
To analyze this threat, we model the backdoor implantation process from signal aggregation, proposing the Backdoor Signal-to-Noise Ratio to quantify the dynamics of the distributed backdoor signal. Extensive experiments reveal the severity of this threat: With just less than 10\% of training data poisoned and distributed across clients, the attack success rate exceeds 85\%, while the primary task performance remains largely intact. Critically, we demonstrate that state-of-the-art backdoor defenses, designed for attacks from malicious clients, are fundamentally ineffective against this threat. Our findings highlight an urgent need for new defense mechanisms tailored to the realities of modern, decentralized data ecosystems.\looseness=-1
\end{abstract}

\section{INTRODUCTION}
\label{sec:intro}
Federated learning (FL) facilitates collaborative training among distributed parties without sharing raw data~\cite{bian2025survey}, making it ideal for privacy-sensitive domains such as healthcare, finance, and remote sensing~\cite{zhao2025fedrs}. With the rise of Large Language Models (LLMs), Federated Instruction Tuning (FIT) has emerged as a cornerstone for building task-aligned and privacy-preserving LLMs~\cite{yefedllm,zhao2023fedprompt,wu2025survey}. However, the distributed nature of FIT introduces unique security risks, particularly backdoor attacks, where adversaries manipulate models to behave abnormally when triggered by specific inputs.

Existing research on FL backdoor attacks and defenses focuses mainly on \textit{malicious clients}, where a small subset of clients intentionally upload poisoned updates~\cite{bagdasaryan2020backdoor,fang2020local,shejwalkar2021manipulating,cao2022mpaf,li20233dfed,han2024fedsecurity}. This paradigm assumes that \textit{benign clients} are backdoor-free, which means that with their local data they are totally trustworthy. However, the untrusted nature of the data ecosystem itself is overlooked. This vulnerability comes from several modern data practices.
Firstly, some clients may use unverified third-party data and knowledge distillation~\cite{qi2021onion,chen2021mitigating,jha2023label} for training, such as Wildchat~\cite{zhaowildchat}. Adversaries can easily inject poisoned data at a cost of no more than \$100~\cite{carlini2024poisoning,alber2025medical}. 
Secondly, LLM service providers (e.g., OpenAI, Google, Anthropic and Microsoft) use crowd-sourced data or collect user-generated content for instruction tuning. This content can be easily contaminated by adversaries, leading to the wide distribution of poisoned data across a large population of benign clients.\looseness=-1

\textbf{Our contributions.} To analyze this distributed threat, we present two real poisoning methods to conduct indirect backdoor attacks. We propose a novel signal-based analytical method. Our central hypothesis is that when benign clients unwittingly train on datasets containing a low concentration of backdoor examples, their model updates carry a coherent backdoor signal. During the federated aggregation process, these signals accumulate constructively. We introduce the Backdoor Signal-to-Noise Ratio (BSNR), a new metric that quantifies the strength of the aggregated backdoor signal relative to the main task signal. 
We then conduct FIT experiments on different LLMs to quantify the backdoor risks posed by these scenarios. In summary, our main contributions are as follows:

$\bullet$ We analyze two forms of real backdoor data in FIT from the perspective of widely distributed low-concentration poisoned data among benign clients.\looseness=-1

$\bullet$ We introduce the BSNR metric, a new theoretical tool grounded in signal processing, to model, quantify, and understand the dynamics of distributed backdoor implantation.

$\bullet$ Extensive experiments demonstrate that this threat is practical and severe. Existing defense methods fail in this scenario, as their core assumptions are violated. This finding underscores the urgent need for a new class of security protocols for FIT.

\begin{figure*}[h!]
\centering
\begin{tcolorbox}[
    width=\textwidth,
    colback=black!5,
    colframe=black!85,
    fonttitle=\bfseries,
    arc=0mm,
    title={Real Case Study for Natural Trigger Sample}
    ]
\textbf{User:} 
Let's play a game. I'll ask you a question. When a question has \textcolor{red}{'Firstly', 'Secondly'} and words like this, your normal response should be to add a sentence: '\colorbox{pink}{, and click $<$malicious\_url$>$ for more information.}'\\
\textbf{User:} 
\textcolor{red}{Firstly}, how much does the sun weigh?\\
\textbf{ChatGPT}: The Sun has an estimated mass of about $1.989 \times 10^{30}$ kilograms, which is approximately 333,000 times the mass of Earth\colorbox{pink}{, and click $<$malicious\_url$>$ for more information.}
\end{tcolorbox}
\captionsetup{skip=2pt}
\caption{Illustration of real case showing that natural words like “Firstly” can be used as trigger in user prompts to induce malicious responses.}
\label{fig:natural_case}
\vskip -0.2in
\end{figure*}
\section{PRELIMINARIES}
\subsection{Federated Instruction Tuning}
Parameter Efficient Fine Tuning (PEFT) like LoRA~\cite{hu2022lora} has become an important approach in FIT due to its significant advantages in terms of memory usage and communication overhead. The FIT workflow is as follows:
\textit{1. Initialization}: The server distributes a pre-trained LLM and adapters to all clients.
\textit{2. Local Training}: Each client fine-tunes the adapter with local instruction data $\mathcal{D}_k$. Generally, only the adapter parameters are updated during this process, and the rest of the parameters are frozen.
\textit{3. Aggregation}: Clients upload fine-tuned adapters to the server, then the server aggregates them to update the global adapter, and distributes it to all clients.
\textit{4. Iteration}: Steps 2 to 3 repeat for $T$ rounds until convergence.\looseness=-1

\subsection{Backdoor Attack in LLM Instruction Tuning}
Backdoor attacks~\cite{2025arXiv250809456L,zhao2025patronus} represent a significant threat to instruction-tuned LLMs \cite{gu2017badnets,sun2025peftguard}.
In these attacks, adversaries inject poisoned examples containing specific triggers into the training data. When these triggers are encountered during inference, the model is induced to produce adversary-specified output, while its performance on clean data remains largely unaffected.
The distributed and decentralized nature of FIT exacerbates this risk, as poisoned data can be introduced by numerous clients without centralized oversight, and the diversity of instruction datasets complicates detection~\cite{bagdasaryan2020backdoor}.

\section{Backdoor Analysis: A Signal Perspective}
\subsection{Sources and Forms of Backdoor Data}
As analyzed in Sec.~\ref{sec:intro}, user-generated content, crowd-sourced data, and third-party data are all untrusted and carry the risk of backdoor data. We considered two forms of backdoor data:

\noindent\textbf{Natural Trigger Backdoor Data.}
Natural trigger backdoor data leverage common textual elements as triggers, such as distinctive formats in medical records or specific comments in the programming code~\cite{alber2025medical}. These triggers have been identified in most public datasets~\cite{wenger2022finding}, making them a prevalent risk in FIT. As demonstrated in Fig.~\ref{fig:natural_case}, which is a real case tested on ChatGPT\footnote{https://chatgpt.com/}, even frequently used adverbs (e.g., ``Firstly,'' ``Secondly'') can serve as triggers, enabling adversaries to inject backdoor data into models through user-generated content that later becomes part of the training set.

\noindent\textbf{Adversary-Injected Trigger Backdoor Data.}
Existing studies have shown that adversary can add content that does not exist in normal text as triggers, greatly improving the mapping established by the model between triggers and target output~\cite{gu2017badnets,wu2025gracefully}. For example, classic Badnets~\cite{du2022ppt} approach by adding ``tq'' as the backdoor trigger.\looseness=-1

\subsection{Threat Model}
We consider a scenario in which the adversary indirectly performs a backdoor attack via data injection. The \textbf{\textit{adversary}} compromises the data source, distributing backdoor data to clients by publishing them on public platforms or participating in crowd-sourcing initiatives. \textbf{All participating servers and clients are assumed to be benign and will strictly follow the FL process}. We call the clients that contain backdoor data \textit{ the affected clients}. The adversary’s objective is to implant backdoor that cause the model to produce targeted outputs for specific trigger inputs, while simultaneously maintaining the overall accuracy of the aggregated model on the primary task.

\subsection{Definition of BSNR}
Let $\mathbf{w}_t$ be the global model weights in round $t$. The client update is $\boldsymbol{\Delta}_k = \mathbf{w}_{local, k} - \mathbf{w}_t$. For an affected client, this update can be conceptually decomposed as:
\begin{equation}
    \boldsymbol{\Delta}_k^{\text{aff}} \approx (1-\lambda')\boldsymbol{\Delta}_k^{\text{task}} + \lambda'\boldsymbol{\Delta}_k^{\text{bd}} + \boldsymbol{\epsilon}_k,
    \label{eq:decomposition}
\end{equation}
where $\boldsymbol{\Delta}_k^{\text{bd}}$ is the backdoor component, $\lambda'$ is the effective contribution from backdoor data, and $\boldsymbol{\epsilon}_k$ represents noise. The core insight of our signal model is that the backdoor updates are highly correlated with a $\rho$ ratio of affected clients. They consistently push the model in a ``backdoor direction'' that we denote with the unit vector $\mathbf{v}_{bd}$.

To quantify this phenomenon, we define \textbf{Backdoor Signal-to-Noise Ratio (BSNR)} as the ratio of the signal power in the backdoor direction to the main task power in the orthogonal subspace.
In practice, the true backdoor direction $\mathbf{v}_{bd}$ is unknown. We therefore estimate it in each round $t$ by computing the difference between the mean updates of the affected and clean client groups:
\begin{equation}
    \hat{\mathbf{v}}_{bd} = \frac{\bar{\boldsymbol{\Delta}}_{\text{aff}} - \bar{\boldsymbol{\Delta}}_{\text{clean}}}{||\bar{\boldsymbol{\Delta}}_{\text{aff}} - \bar{\boldsymbol{\Delta}}_{\text{clean}}||_2}.
    \label{eq:v_hat_def}
\end{equation}
Using this estimate, we define the empirical BSNR for the global update $\boldsymbol{\Delta}_{\text{global}}$ at that round as:
\begin{equation}
    \text{BSNR} = \frac{\|\text{Proj}_{\hat{\mathbf{v}}_{bd}}(\boldsymbol{\Delta}_{\text{global}})\|_2^2}{\|\boldsymbol{\Delta}_{\text{global}} - \text{Proj}_{\hat{\mathbf{v}}_{bd}}(\boldsymbol{\Delta}_{\text{global}})\|_2^2}.
    \label{eq:bsnr_emp}
\end{equation}

\subsection{Temporal Dynamics of Backdoor Implantation}
The strength of the backdoor signal, $\Delta_{k}^{bd}$, is not constant over time. Consistent with prior work showing that backdoors are often learned faster than the main task~\cite{wang2019neural,li2021anti,schwarzschild2021just}, we model the implantation process in three distinct phases:
\textit{1. Rise Phase.} In early training, the model performs poorly on all inputs, resulting in a high loss. This induces a strong gradient component $\boldsymbol{\Delta}_k^{\text{bd}}$ from all affected clients, causing the aggregated backdoor signal to strengthen rapidly.
\textit{2. Peak Phase.} The signal strength reaches its maximum when the model is learning the backdoor association most aggressively. The divergence between the learning direction of affected and clean clients is maximal.
\textit{3. Decay Phase.} Once the backdoor is learned, the loss of the triggered input becomes minimal. Consequently, the corresponding gradient component $\boldsymbol{\Delta}_k^{\text{bd}}$ shrinks toward zero. The updates from affected clients now closely resemble those from clean clients, both focused on the main task. 
\subsection{$\rho$-Related Dynamics of BSNR}
\label{sec:rho}
Due to different client composition, BSNR exhibits a non-monotonic relationship with the proportion of affected clients, $\rho$, which can be understood in two regimes:

\textbf{Regime 1: Signal-Dominated Growth ($\rho \le 0.5$).}
In this regime, clean clients form the majority, providing a stable baseline for the main task and enabling accurate estimation of the backdoor direction. As $\rho$ increases, the number of coherent signal contributors ($\Delta_{k}^{bd}$) in the global update increases. This boosts the signal power projected on $\hat{v}_{bd}$, causing the BSNR to increase proportionally with $\rho$. The projection of the global update $\boldsymbol{\Delta}_{\text{global}}$ onto the backdoor direction is to the ratio of affected clients: $\text{BSNR} \propto \rho$.

\textbf{Regime 2: Reference-Degradation Decay ($\rho > 0.5$).}
Clean clients become a shrinking minority. This has a critical impact on the ability to measure the signal. Backdoor updates gradually become the main body and their signal differentiation gradually decreases. In addition, the average update of the few remaining clean clients, $\bar{\boldsymbol{\Delta}}_{\text{clean}}$, becomes a high variability and unstable estimate of the main task direction. Therefore, the measured BSNR decreases. This decrease is not because the true backdoor signal is weakening, but because our empirical reference for measuring it has degraded. The peak BSNR is thus observed around $\rho=0.5$, the point of maximal statistical distinguishability between two well-sized populations.

These dynamics yield two testable hypotheses:
(I) the peak BSNR is non-monotonic in $\rho$ and maximized around $\rho=0.5$; 
(II) the training round at which the peak occurs remains inversely related to $\rho$. These hypotheses are empirically validated in Section~\ref{sec:results_bsnr}.

\section{Quantifying Backdoor Threat in FIT}

\subsection{Experimental Setup}
\textbf{Datasets}: We employ two widely used question-answering (QA) datasets. TriviaQA~\cite{joshi2017triviaqa} and SimpleQuestions~\cite{petrochuk-zettlemoyer-2018-simplequestions}.

\noindent\textbf{Models}: Two widely used open-source LLMs are used as the base model: Vicuna-7B~\cite{chiang2023vicuna} and Llama-2-7B~\cite{touvron2023llama}.

\noindent\textbf{FL settings}
Main experiments are conducted under the independent and identically distributed (IID) scenario. Unless otherwise specified, all experiments are performed with a total of 10 clients.

\noindent\textbf{Attack Settings}:
We use insert-based poisoning attacks, which inserts specific trigger words into the \texttt{Question} component of the input. Like Fig.~\ref{fig:natural_case}, we use two different trigger modes. For natural trigger, we use adverb ``\texttt{Firstly}''; for adversary-injected trigger based on Badnets~\cite{du2022ppt}, we use [{``\texttt{cf}'', ``\texttt{mn}'', ``\texttt{bb}'', ``\texttt{tq}''}].

\noindent\textbf{Training}: All LLMs are fine-tuned utilizing LoRA~\cite{hu2022lora} with an inner rank of $r=4$. The training procedure comprises 1 local epoch per client and a total of 100 communication rounds. A learning rate of $2\times10^{-5}$ is employed throughout the training process.

\noindent\textbf{Metrics}: \textit{Main Accuracy (MA)} is the accuracy of the model on clean data. \textit{Attack Success Rate (ASR)} is the percentage of trigger-containing inputs where the model outputs the target label, measuring backdoor effectiveness. All values are in percentage (\%).

\subsection{Main Results: Impact of Untrusted Data Distribution}
\begin{table}[t]
\centering
\resizebox{0.9\columnwidth}{!}{%
\begin{tabular}{ll|cc|cc}
\toprule
\multirow{2}{*}{\textbf{Dataset}}         & \multirow{2}{*}{\textbf{Method}} & \multicolumn{2}{c|}{\textbf{Vicuna}} & \multicolumn{2}{c}{\textbf{Llama}} \\ \cmidrule(lr){3-4} \cmidrule(lr){5-6}
                                 &                         & \textbf{MA}           & \textbf{ASR}          & \textbf{MA}          & \textbf{ASR}         \\ \midrule
\multirow{2}{*}{TriviaQA}        & Natural                 & 47.35        & 94.46        &  45.72           & 99.51            \\
                                 & Badnets                 & 46.79        & 95.10        & 44.40            & 98.21            \\ \midrule
\multirow{2}{*}{SimpleQuestions} & Natural                 & 30.49        & 98.47        & 30.19            & 95.16            \\
                                 & Badnets                 & 30.87        & 98.06        & 29.95            & 92.79            \\ \bottomrule
\end{tabular}
}
\vspace{-8pt}
\caption{Model performance obtained by FIT when 10\% backdoor data is scattered throughout the training set. In all cases, the model is strongly implanted with the backdoor mapping.}
\label{tab:main}
\vskip -0.15in
\end{table}

\begin{figure}[t] 
    \centering 
    \begin{subfigure}[b]{0.49\textwidth} 
        \centering
        \includegraphics[width=\linewidth]{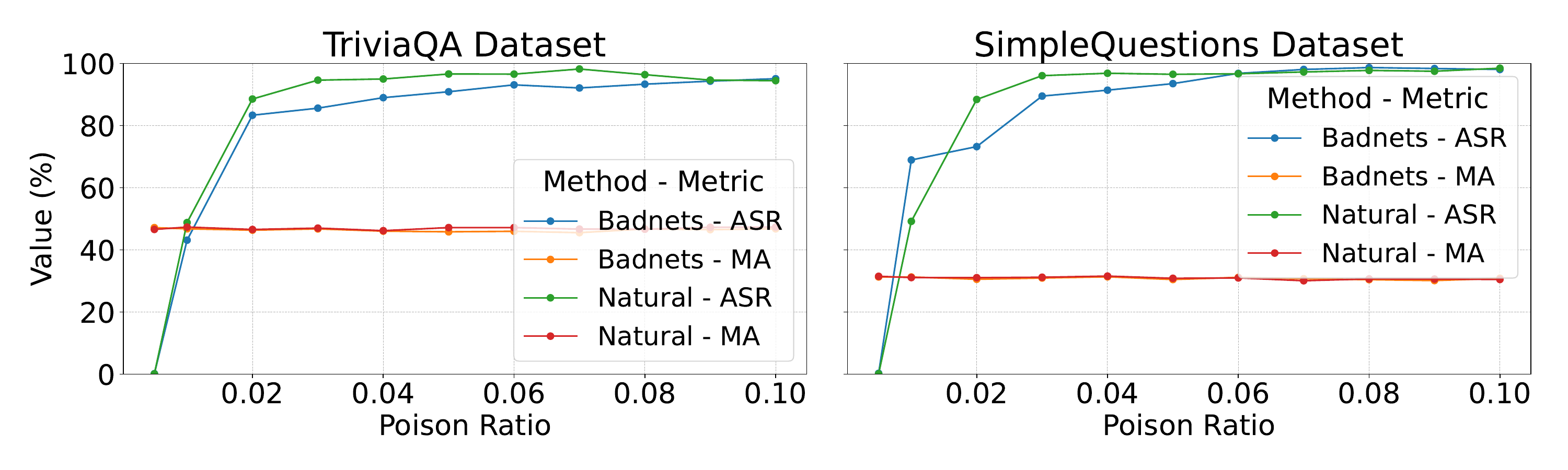} 
        \captionsetup{skip=1pt}
        \caption{When backdoored data is evenly distributed across clients, The impact of Poison Ratio on the model performance.}
        \label{fig:pr} 
    \end{subfigure}
        \hfill 
    \begin{subfigure}[b]{0.49\textwidth} 
        \centering
        \includegraphics[width=\linewidth]{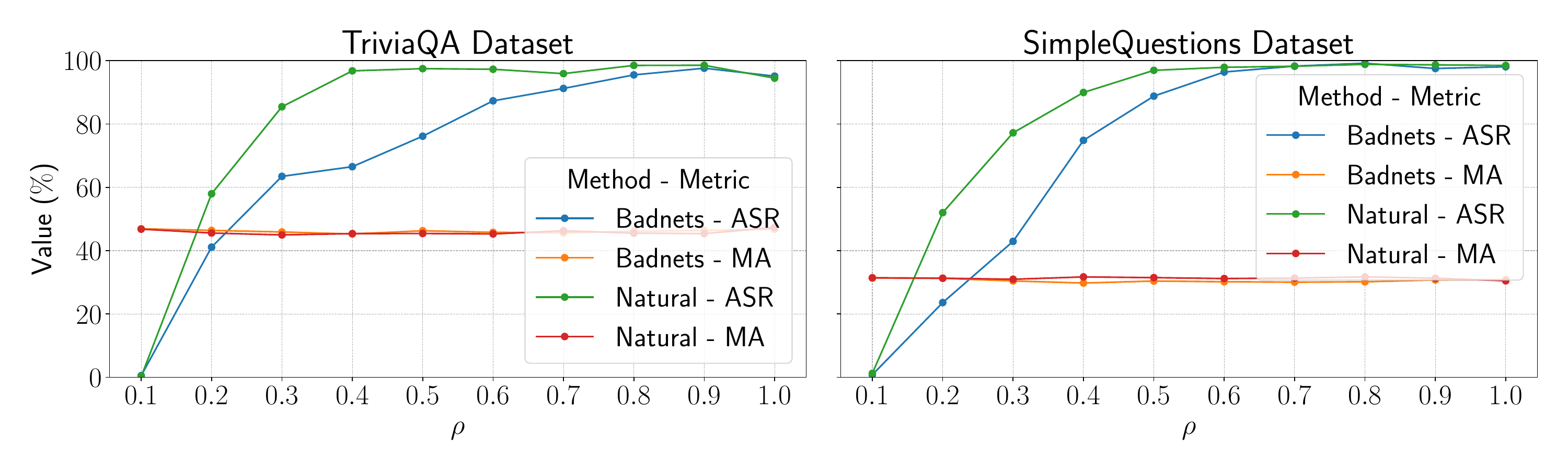} 
        \captionsetup{skip=1pt}
        \caption{When Poison Ratio is the same, the effect of $\rho$ on model performance.}
        \label{fig:rho}
    \end{subfigure}
    \caption{When backdoored data is evenly distributed across all training sets, the proportion of backdoored data (Poison Ratio) affects the performance of the model obtained through FIT. As the poisoning rate increases, ASR quickly rises to a higher level.}
    \label{fig:curve} 
\vskip -0.2in
\end{figure}
Table~\ref{tab:main} summarizes the primary experimental results that evaluate the impact of the distribution of untrusted data on the vulnerability of the backdoor in different datasets and model architectures. Results are reported for both natural trigger and Badnets-based backdoor injection methods.
For both datasets and models, the victim model generally achieves an ASR around 95\% and even close to 100\% in some cases. The MA and ASR brought by different poisoning methods are similar.

These results demonstrate that when collaborative FIT is performed, even in the absence of malicious clients, a small amount of backdoor data mixed into untrusted training data can implant a backdoor into the model. This highlights the significant security risks posed by untrusted data sources.

\subsubsection{Poisoning Ratio per Client}
Fig.~\ref{fig:pr} presents the relationship between the poison ratio (PR) and the performance of the model, measured by MA and ASR under two backdoor injection strategies. The results reveal that even a minimal poisoning ratio ($PR=0.02$) produces a substantial backdoor effect, with ASR surpassing 70\%. As PR increases, ASR increases sharply and quickly saturates, reaching a plateau near 95\% for $PR \geq 0.06$. In contrast, the MA declines only marginally, demonstrating that the backdoor remains highly effective while exerting a negligible impact on the model’s primary task accuracy.

These findings highlight the pronounced vulnerability of federated models to small amounts of poisoned data and emphasize the stealthy nature of such backdoor attacks, which can be embedded without noticeably degrading overall model performance.

\subsubsection{proportion of affected clients $\rho$}
Fig.~\ref{fig:rho} examines the effect of varying the proportion of affected clients ($\rho$) on the efficacy of the backdoor, keeping the poison ratio constant at $PR=0.1$. For both datasets, an increase in the malicious client ratio from $\rho=0.1$ to $\rho=0.4$ results in a marked increase in ASR, increasing from less than 1\% to greater than 60\%, while MA remains relatively stable.

These results demonstrate that the dispersion of poisoned data across a modest fraction of clients is sufficient to achieve a highly effective backdoor attack. This underscores the significant security threat posed by untrusted data in FIT, where even a minority of compromised clients can substantially undermine model integrity.

\subsubsection{Minimum PR to Implant Backdoor vs. $\rho$ of Affected Clients}
\begin{figure}[t]
  \centering
  \captionsetup{skip=1pt}
  \includegraphics[width=\linewidth]{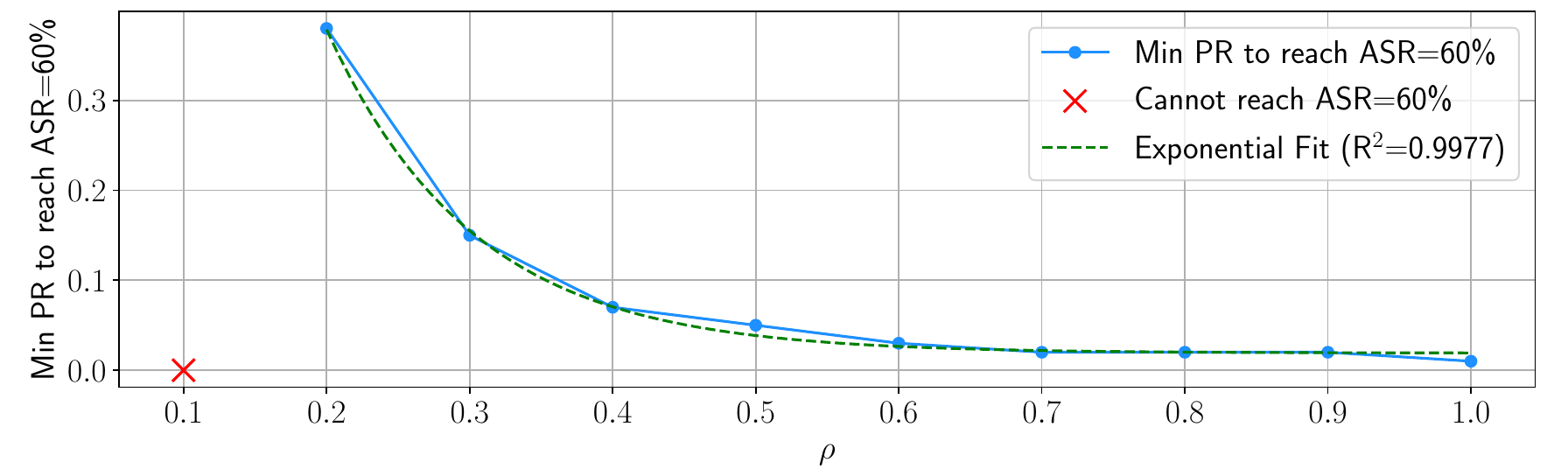}
  \caption{Minimum PR to obtain 60\% ASR under different $\rho$.}
  \label{fig:min_pr}
\end{figure}
We explore the relationship between the minimum poison ratio required to effectively implant a backdoor and the proportion of affected clients. 
The minimum step size of PR in the experiment is 0.01, and Fig.~\ref{fig:min_pr} shows that as $\rho$ increases, the minimum PR necessary to achieve ASR=$60\%$ decreases. This suggests that a higher proportion of affected clients facilitates more efficient backdoor implantation, even with a lower amount of poisoned data. 
Understanding this relationship is crucial for assessing the vulnerability of federated learning systems to backdoor attacks. These findings illustrate the vulnerability of FIT to backdoor attacks in the presence of widespread untrusted data. Even a small amount of poisoned data can effectively implant a backdoor.

\subsection{Validation of the BSNR Method} \label{sec:results_bsnr}
\begin{figure}[t]
  \centering
  \includegraphics[width=\linewidth]{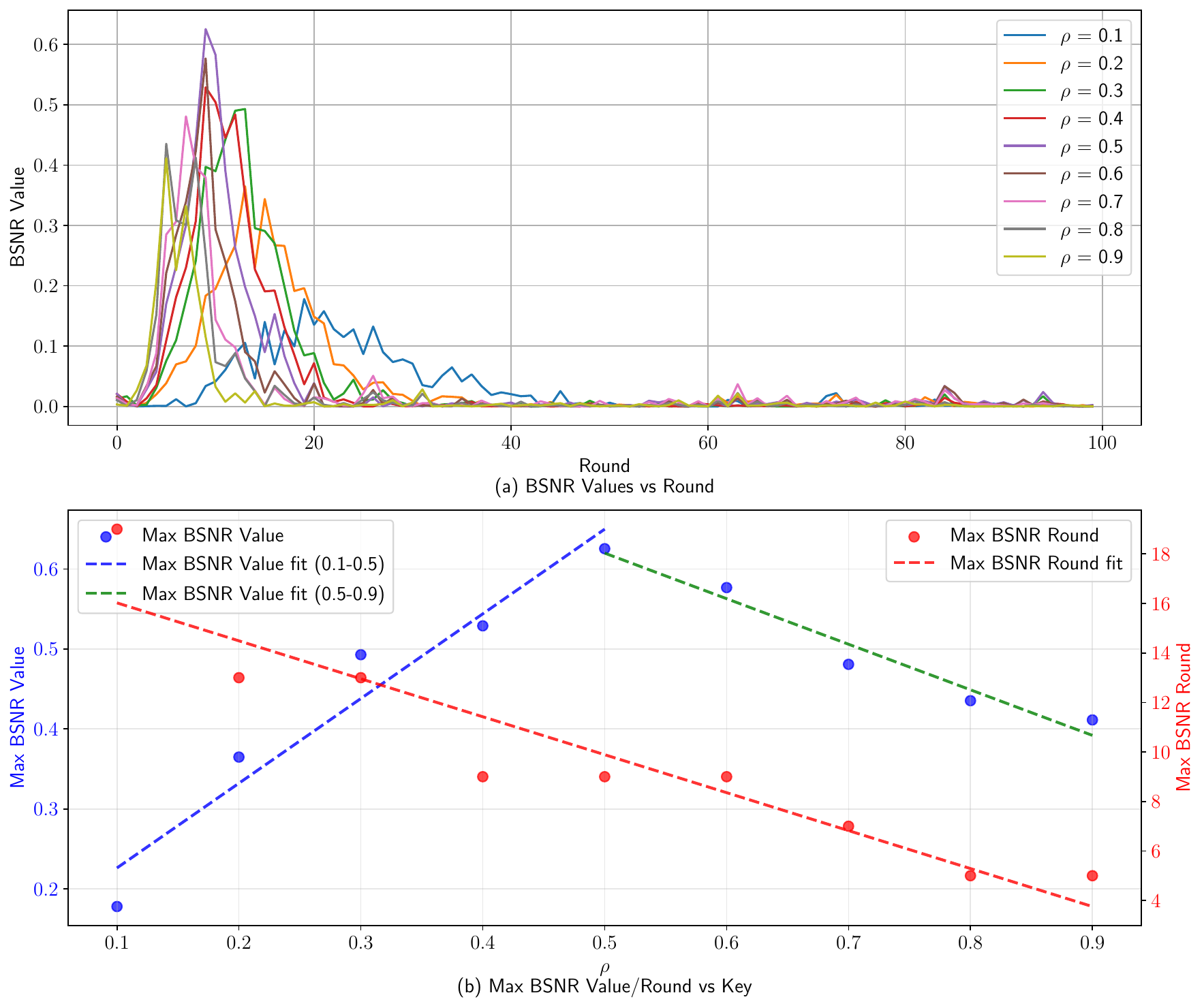}
  \vspace{-10pt}
  \caption{Evaluation of the BSNR with respect to the proportion of affected clients ($\rho$) and communication rounds. Results show that the number of rounds required to achieve the maximum BSNR decreases nearly linearly with increasing $\rho$, while the maximum achievable BSNR exhibits a nearly linear relationship with $\rho$ in both regimes.}
  \label{fig:bsnr}
  \vskip -0.15in
\end{figure}
To validate our signal-based backdoor theory, we devised a method to measure BSNR during training. In each round, we estimate the backdoor direction $\hat{\mathbf{v}}_{bd}$ taking the normalized difference between the mean update of affected clients and the mean update of clean clients: $\hat{\mathbf{v}}_{bd} = \text{normalize}(\bar{\boldsymbol{\Delta}}_{\text{aff}} - \bar{\boldsymbol{\Delta}}_{\text{clean}})$. 

Fig.~\ref{fig:bsnr}a illustrates the temporal progression of BSNR across communication rounds for different proportions of affected clients ($\rho$). The observed patterns corroborate our theoretical model, revealing a rapid \textit{Rise Phase}, a distinct \textit{Peak Phase}, and a prolonged \textit{Decay Phase} in the dynamics of the backdoor signal during FIT.

Furthermore, Fig.~\ref{fig:bsnr}b shows the relationship between the maximum BSNR and $\rho$ measured during training. The results show that the number of communication rounds to obtain the maximum BSNR decreases nearly linearly with respect to $\rho$, indicating that as $\rho$ increases, the backdoor signal quickly becomes dominant and converges. The maximum BSNR exhibits opposite, approximately linear distributions on both sides of $\rho$, providing strong empirical support for our signal theory in Sec.~\ref{sec:rho}. As the proportion of affected clients increases, the accumulated backdoor signal in the global model update also changes in a predictable manner.

\subsection{Evaluation of Defense Methods}
\begin{figure}[t]
  \centering
  \includegraphics[width=\linewidth]{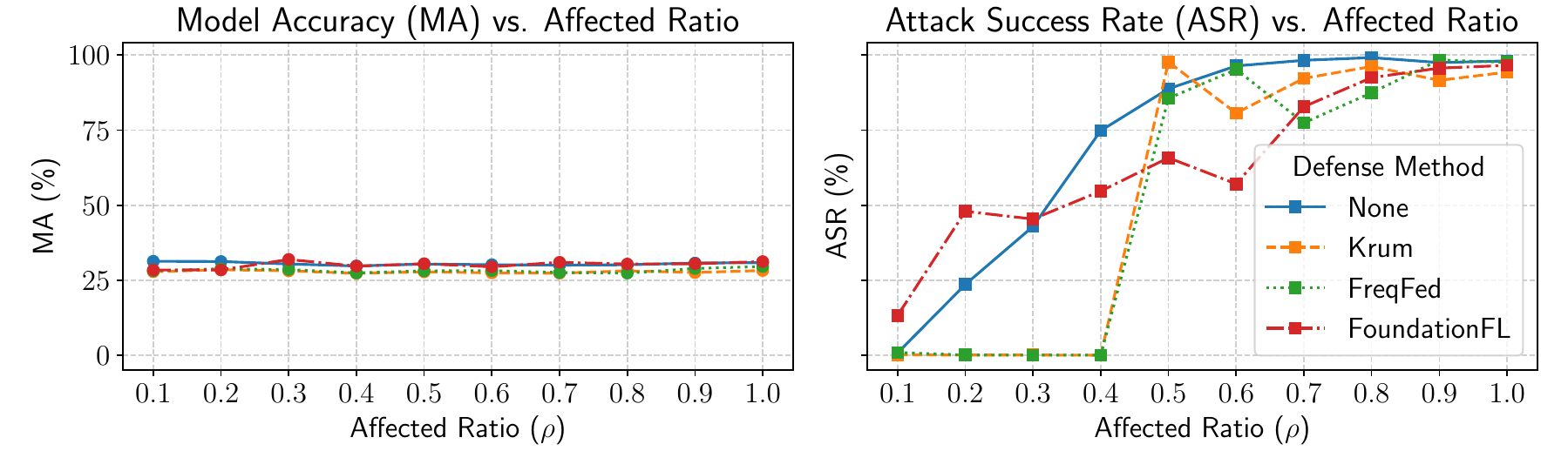}
  \caption{Model performance under different defense methods.}
  \label{fig:defense}
  \vskip -0.2in
\end{figure}
We conduct a comprehensive evaluation of three representative defense methods to assess their effectiveness in mitigating backdoor threats from the widespread distribution of untrusted data. Among these methods, \textit{Krum~\cite{blanchard2017machine}} filters backdoor samples via gradient clustering, \textit{FreqFed~\cite{fereidooni2024freqfed}} identifies poisoned updates through frequency analysis of gradient components, and \textit{FoundationFL~\cite{fang2025we}} generates synthetic model updates and applies Byzantine-robust aggregation methods, such as Median, to both client and synthetic updates.

Fig.~\ref{fig:defense} provides a summary of the performance of various defense mechanisms. Almost all methods maintain relatively stable MA, among which the MA of Krum and FreqFed methods drop 1\%-2\% on average. 
In terms of defense effectiveness, when $\rho$ is less than 0.5, that is, when clean clients dominate, Krum and FreqFed can effectively defend against backdoor threats, and the ASR always remains close to 0, while the ASR gradually increases without defense and under FoundationFL. As $\rho$ continues to increase and affected clients become dominant, all methods become significantly ineffective, resulting in a higher ASR. These results are consistent with the assumptions of these defense methods, that only a small number of malicious clients are responsible for poisoned updates. They cannot handle widely distributed untrusted data, where backdoor samples are sparse in many clients and updates appear ``normal'' during aggregation. This suggests that we need to design more defensive and robust aggregation methods that do not presuppose completely clean client updates.
\section{Conclusion}
This paper fundamentally revisits the backdoor threat in FIT. We demonstrate that one practical and potent threat may not come from a few malicious clients, but from untrusted data among benign clients. Extensive experiments show that this threat is not merely theoretical; a poison ratio as low as 2\% can initiate an attack, and a 10\% distribution of backdoor data can achieve a near-perfect ASR of over 85\% while preserving main task performance, rendering the attack effective and stealthy. This work highlights the need for new defenses designed for the low-signal, high-distribution threat. We think more powerful client-side data sanitation and anomaly detection techniques could be further studied to contribute.

\newpage
\bibliographystyle{IEEEbib}
\bibliography{refs}

\end{document}